 \documentclass[preprint]{aastex}

\received{8 March 2001}
\accepted{17 April 2001}

\shorttitle{86~GHz VLBP of 3C\,273 and 3C\,279}
\shortauthors{Attridge}

\begin{document}


\title{86~GHz Very Long Baseline Polarimetry of 3C\,273 and 3C\,279 \\
    with the Coordinated Millimeter VLBI Array}

\author{Joanne M.~Attridge}
\affil{MIT Haystack Observatory, Off Route 40, Westford, MA 01886}
\email{jattridge@haystack.mit.edu}

\begin{abstract}
86~GHz Very Long Baseline Polarimetry probes magnetic field structures
within the cores of Active Galactic Nuclei at higher angular resolutions
and a spectral octave higher than previously achievable. Observations of 
3C\,273 and 3C\,279 taken in April 2000 with the Coordinated 
Millimeter VLBI Array have resulted in the first total 
intensity (Stokes $I$) and linear polarization VLBI images 
reported of any source at 86~GHz. These results reveal the 86~GHz
electric vector position angles within the jets of 3C\,273 and 
3C\,279 to be orthogonal to each other, and the core of 3C\,273 to be
unpolarized. If this lack of polarization is due to Faraday depolarization alone, the 
dispersion in rotation measure is $\gtrsim$90000~rad m$^{-2}$ for 
the core of 3C\,273. 
\end{abstract}



\section{Introduction}
Magnetic fields trace the flows of relativistic plasma ejected from the
cores of Active Galactic Nuclei (AGN) in the form of powerful jets. 
Ordering of the jet magnetic
fields by shocks (e.g.~Hughes, Aller, \& Aller 1985, Marscher \& Gear 1985) or
shear \citep{war94} occurs within a fraction of a parsec (pc) from the jets'
points of origin. 
Though synchrotron theory predicts 
sources may exhibit up to 75\% fractional linear polarization 
($m$) in the optically thin regime, this level of polarization has never
been observed in AGN cores. It is likely that the modest $m$ observed in AGN 
cores at centimeter wavelengths results from disordered magnetic fields, 
depolarization, and high optical depths. 86~GHz ($\lambda3.5$mm) Very Long 
Baseline Polarimetry (VLBP) should be sensitive
to intrinsic field order and may reveal the first instances of magnetic 
field ordering mechanisms, for the effects of Faraday rotation ($\propto\lambda^2$) 
and depolarization are reduced at high frequencies. 

\section{Observations and Calibration}
Images of 3C\,273 and 3C\,279 were produced from data taken with four
antennas (Fort Davis, Los Alamos, Pie Town, Kitt Peak 12-m)
during the Coordinated Millimeter VLBI Array (CMVA)\footnote{CMVA operations 
and correlation at Haystack 
are conducted under support from the National Science Foundation.} 
program CA01 over $\sim$7~hours on 2000 April 17 (epoch 2000.30).
The data were recorded at 86~GHz with a total bandwidth of 64~MHz. 
This four station subset of the complete CMVA run was correlated on the 
NRAO\footnote{The National Radio Astronomy Observatory is a facility of 
the National Science Foundation, operated under a cooperative agreement by 
Associated Universities, Inc.} correlator as a certification test of the 
Haystack Mark IV correlator system. Approximately $2^{h} 45^{m}$  
total integration were devoted to each source leading to a theoretical RMS 
noise of $\sim6$~mJy beam$^{-1}$. The complete, eight station, $\sim$24 
hour CMVA run, providing full tracks on both 3C\,273 and 3C\,279 
with a global array of mixed single- and dual-polarization antennas, will 
be correlated at Haystack later in 2001.

Calibration and hybrid imaging were done using the NRAO AIPS
and Caltech DIFMAP packages. Total intensity is notated as $I$, and
linear polarization is represented by the complex quantity 
$P=Q+iU=pe^{2i\chi}=mIe^{2i\chi}$, where $p=(Q^{2}+U^{2})^{1/2}=mI$ is
the polarized intensity, and $\chi$ is the 
electric vector position angle (EVPA). Explanations of VLBP 
calibration and imaging may be found in \citet{cot93}, 
Roberts, Wardle, \& Brown (1994), and \citet{att99a}.

The calibration process for 86~GHz VLBI data must account for 
complexities not present in lower frequency data \citep{att99b}.
For example, coherence times at 86~GHz are typically only 
$\sim10$~seconds, severely limiting choice of solution interval in 
many AIPS tasks. Many antennas succumb to pointing 
problems at 86~GHz, primarily due to wind-loading. The impact of 
these problems escalates when considering that 86~GHz data is lower 
(by a factor of $\gtrsim2$) in SNR than 43~GHz data.

Of primary concern was calibration of the instrumental polarization
(D-terms). Careful tests were performed to determine the
veracity of the results. First, the sources were imaged with no D-term
solution applied. Resulting images have scattered polarization
throughout the fields of both 3C\,273 and 3C\,279. Next, the D-terms were
calculated individually for each source using the AIPS task LPCAL,
applied to the data, and then the data were imaged. The maps presented
in this paper are derived from these individual applications of LPCAL.
As a test of the LPCAL
results, the D-terms for 3C\,273 were then applied to 3C\,279 and vice
versa with no apparent change in the maps. The task PCAL was run on
3C\,279\footnote{PCAL algorithms are restricted to very simple source
structure, and therefore could not be applied to the extended structure
of 3C\,273.} and the resulting D-terms were applied to both sources,
again with no change in the maps.

For 3C\,273, the D-terms range from $\sim3-16\%$, and for
3C\,279, from $\sim2-21\%$, resulting in a typical value of $12\%$.
These ranges represent the same set of D-terms offset by a
vector constant, arising from the special case of the four antennas
having effectively the same parallactic angle coverage \citep{con69,rob94}.

\section{Results}
Naturally weighted images in $I$ and $P$ are presented in 
Figures~1 and 2. These images are the first 86~GHz VLBI Stokes $I$ 
and $P$ images presented for any source. 
Currently, there is no known calibrator by which to register absolute 
EVPAs at 86~GHz; therefore sticks in the $P$ images represent the 
orientation of the electric field vectors assigned an arbitrary 
instrumental position angle, rotated through an adopted 90$^{\circ}$ 
EVPA calibration constant as explained in \S 4.2. Note that the 
EVPAs of the two sources {\em relative to each other} are correctly 
registered.  

The $I$ images presented in Figures 1$a$ and 2$a$ have dynamic 
ranges of $\sim$100:1, and the Fourier transforms of the clean components
represented by the images result in the best matches to the $(u,v)$ data 
among many trials of hybrid mapping and self-calibration, down to time
scales of thirty seconds in phase, and two minutes in amplitude. 
DIFMAP's ``modelfit'' routine was used to fit 
individual components in $I$, $Q$, and $U$. Because $Q$ and $U$ 
positions were restricted to the locations of the $I$ peaks, 
resultant values of $m_{86}$ do not necessarily represent the true
maximum $m_{86}$ of a component. Model fits for both 3C\,273 and 
3C\,279 are presented in Table 1. 

\subsection{3C\,273}
3C\,273 (J1229+0203) is a low optical polarization quasar (LPQ) with $z=0.158$
\citep{str92},
corresponding to $2.63 h^{-1}$~pc/mas.\footnote{$H_0 = 70
h$~km~s$^{-1}$~Mpc$^{-1}$ and $q_0 = 0.05$ are assumed in all calculations.}
A typical core-jet source, 3C\,273 contains a bright component 
to the east which is resolved in the north-south direction, and a jet 
extending to the west-southwest. The $P$ image shown in Figure~1$b$ displays
linear polarization only in the jet component (C2) adjacent to the core 
(D); the core itself is unpolarized to a limit of $1\%$. Model fits to the data  
(Table~1) show that C2 has $m_{86}\sim11\%$.

Historically, at frequencies lower than 86~GHz 3C\,273 displays very 
low levels of fractional core
polarization with increasing levels further down the jet. 3C\,273's
variable nature and the differing resolutions provided by images at
other frequencies, make temporal and spatial comparisons of all but
the core quite difficult. The most recent, published VLBI values of 
$m$ for 3C\,273 are presented by \citet{tay98} and \citet{lis00} 
(hereafter L\&S). At 15~GHz (epoch 1997.07), 3C\,273 displayed 
$m_{15}\sim0.2\%$ in the core \citep{tay98}. At 22~GHz (epoch 1999.03), 
L\&S found a maximum core polarization of $m_{22}<0.1\%$; 43~GHz results, 
also at epoch 1999.03, reveal a maximum core polarization of $m_{43}<0.6\%$.

\subsection{3C\,279}
3C\,279 (J1256-0547) is a compact source with a jet extending to the
west-southwest. It is a high optical polarization quasar (HPQ) with
$z=0.536$ \citep{mar96}, corresponding to $5.89 h^{-1}$~pc/mas.\footnotemark[4] 
In the $P$ image of Figure~2$b$, 
both the core (D) and the jet component (C1) are linearly polarized. In 
fact, C1 is highly polarized, displaying $m_{86}\sim20\%$, while 
the core displays $m_{86}\sim5\%$ (Table~1).  

The 3C\,279 results presented here are consistent with past high
frequency VLBI observations of the source. Taylor (1998, 2000) found
15~GHz fractional polarization of $m_{15}=4.0\%$ in the core
of 3C\,279 at epoch 1997.07, and $m_{15}=7.6\%$ at epoch 1998.59 - 
an increase of 90\%.\footnote{3C\,279 experienced a significant outburst at 
37~GHz \citep{weh00} and 90~GHz in late-1997 (M. Tornikoski 2000, 
private communication), and at 15~GHz in mid-1998
(\url{http://www.astro.lsa.umich.edu/obs/radiotel/umrao.h-tml}).} 
L\&S found upper limits on the core polarizations to be 3.9\% at 22~GHz, 
and 6.6\% at 43~GHz; both at epoch 1999.03. Epoch 1997.58 VLBA 
observations at 43~GHz show $m_{43}=4.5\%$ in the core of
3C\,279 (A.~P. Marscher et al. 2001, in preparation).

\section{Discussion}

\subsection{Proper Motion Extrapolation}
Using an average proper motion of 1~mas/yr for 3C\,273 as derived 
from six epochs of dual-frequency data by \citet{hom01}, extrapolation 
of jet components from epoch 1999.03 (L\&S) to the 86~GHz epoch 
of 2000.30 was performed. L\&S's component H evolves to a distance of
2.1~mas from the core, coincident with 86~GHz component C1. Two 
intermediate components from L\&S, i and j, are located at 1.4 and 
1.6~mas, respectively, from the core. These two components lie between
components C1 and C2 of the 86~GHz data.

Similarly, a proper motion of 0.2~mas/yr, also from \citet{hom01},
was used to extrapolate L\&S's epoch 1999.03 3C\,279 jet components to 
epoch 2000.30. After extrapolation, L\&S's component F falls 
at a distance of 0.9~mas from the core,
consistent with 86~GHz component C1. Component F/C1 reveals high levels
of polarization at 22~GHz ($m_{22}=18.3\%$), 43~GHz ($m_{43}=25.7\%$) 
and 86~GHz ($m_{86}\cong20\%$). High levels of fractional polarization 
($\geq20\%$) are not unique, and have been seen in a number of sources; 
e.g.~32\% in OJ\,287 \citep{gab01}, 20\% in BL~Lacertae \citep{den00}, 
and 29\% in 3C\,454.3 \citep{caw96}.
Note that the peak $m_{86}\sim26\%$ in 3C\,279 
is located 0.85~mas from the core; between the core and C1.

\subsection{Electric Vector Position Angles}
As previously stated, no absolute EVPA ($\chi$) calibration can yet be
performed on the 86~GHz data. Nonetheless, the EVPAs of the two
sources may be compared to each other, and to values measured at
other frequencies. 

If $\Delta\chi=\chi_{3C\,273}-\chi_{3C\,279}$, then 
$\Delta\chi_{86}=63^{\circ}$ is the difference between the EVPAs of 
3C\,273 and 3C\,279 at 86~GHz. Similarly, single dish data from
the James Clerk Maxwell Telescope (JCMT) results in 
$\Delta\chi_{270}=87^{\circ}$ at 270~GHz, epoch 1995.98 \citep{nar98}, 
and $\Delta\chi_{350}=97^{\circ}$ at 350~GHz, epoch 2000.26 (J.~A. Stevens
2001, private communication).
Epoch 1999.03 VLBI observations at 43~GHz show $\Delta\chi_{43}=60^{\circ}$ 
(L\&S). The 86~GHz results are thus consistent with $\Delta\chi$ values at
adjacent frequencies. Comparing the $\chi$ values of 
3C\,273 and 3C\,279 at 86~GHz with those at 43~GHz suggests an offset
between the two frequencies of $\sim90^{\circ}$. This comparison is reasonable,
as L\&S's 43~GHz data is closest in resolution to and is
only one spectral octave off from the 86~GHz data. It would therefore
be reasonable to assume that the proper EVPA calibration constant
at 86~GHz is $\sim90^{\circ}$. This estimated 90$^{\circ}$ calibration
constant has been applied to the $P$ images shown in Figures 1$b$ and 2$b$.

It is well known that the EVPAs of AGN jets are closely related to
their jet position angles, $\theta$ (e.g.~Cawthorne et al. 1993).
VLBI observations at many frequencies generally show the EVPA of 3C\,273 
to be transverse ($E_{\perp}$) to the VLBI jet axis, and the EVPA of 
3C\,279 to be longitudinal ($E_{\parallel}$) to the VLBI jet axis 
(e.g.~15~GHz: Homan \& Wardle 1999; 22 and 43~GHz: L\&S; 43~GHz: A.~P. 
Marscher et al. 2001, in preparation). Aligning the EVPAs measured in single 
dish measurements at 270~GHz \citep{nar98} and 350~GHz (J.~A. Stevens 
2001, private communication) with the VLBI jet position angle of a lower 
frequency produces the same results: 3C\,273 displays $E_{\perp}$ and 
3C\,279 displays $E_{\parallel}$. The 86~GHz $P$ images presented in Figures
1$b$ and 2$b$ reveal 3C\,273's EVPA to be transverse to the
jet direction, and 3C\,273's to be longitudinal to the jet
direction, consistent with results at adjacent frequencies. This suggests
that application of the estimated 90$^{\circ}$ EVPA calibration constant 
suggested above is appropriate.
 
\subsection{Rotation Measures and Faraday Screens}
Taylor (1998, 2000) and L\&S derive absolute Faraday rotation measures (RMs) in excess of 
1000~rad~m$^{-2}$ in the cores of many quasars, including 3C\,273 and 3C\,279. 
Beyond projected distances of $\sim20$ parsecs from the core, RMs in
the jets of quasars drop to $\left|RM\right|<100$~rad m$^{-2}$\citep{tay98}.
The presence of large RMs suggests that parsec-sized Faraday screens
with organized magnetic fields will be found near the cores of quasars (L\&S).

As Faraday rotation is $\propto\lambda^{2}$, RMs are expected to diminish
at high frequencies. The observations presented here challenge this 
expectation by showing the cores of 3C\,273 and 3C\,279 to have low levels
of polarization even at 86~GHz. The core of 3C\,273 is unpolarized 
in VLBP observations up through 43~GHz, and the addition of the 86~GHz results 
($m_{86}<0.6\%$) show that this trend continues at higher frequencies. 
It is interesting to explore the case where depolarization of 3C\,273's core
is attributed to Faraday depolarization alone. Assuming that $\sigma_{RM}$ is 
the standard deviation of a Gaussian distribution of many RMs over a finely spaced 
grid \citep{bur66}, and that $m_{86}\sim11\%$ of the jet represents a purely
undepolarized state, $\sigma_{RM}$ must be $\gtrsim$90000~rad m$^{-2}$ for 
the core of 3C\,273. 

Alternatively, the magnetic field in the core of 3C\,273 may be 
initially tangled, becoming ordered somewhere between the core (D) and the
adjacent polarized jet component (C2). The degree of order of the magnetic 
field may be calculated assuming the magnetic field in 
the unshocked jet fluid contains a small-scale tangled component plus a uniform
component aligned perpendicular to the line of sight. The resulting
fraction of magnetic energy in the tangled component in 3C\,273 is then 
$\sim89\%$ \citep{bur66,war94}.

\section{Conclusions}
The CMVA results presented here validate 86~GHz VLBP as a practical and potent
new tool to probe magnetic field properties of AGN cores. 
With only 86~GHz observations, it is impossible 
to distinguish between the depolarizing effects
of a Faraday screen, and a tangled magnetic field. Future simultaneous, multi-frequency
VLBP observations, including ones at 86~GHz, are crucial in testing 
various depolarization mechanisms in the cores of 3C\,273 and 3C\,279.
With the addition of new 86~GHz antennas, particularly those that provide
long baselines, resolutions much higher than those currently attainable 
at 43~GHz should be achieved. High resolution observations should 
set limits on the sizes of Faraday screens in both 3C\,273 and 3C\,279 
by resolving the cores of these objects into smaller components,
thereby testing the discussion of \S 4.3.


\acknowledgments
J.~M.~A. is supported by the MIT-CfA Postdoctoral Fellowship at Haystack.
Radio astronomy at Haystack is supported by the NSF through the
Northeast Radio Observatory Corporation under grant AST-9727353. 
This research has made use of data from the University of Michigan Radio 
Astronomy Observatory which is supported by funds from the University 
of Michigan.

The author thanks the referee, A.~P. Marscher, for his kind words and
thorough reading of this manuscript; 
D.~C. Homan, R.~B. Phillips, and J.~F.~C. Wardle for 
rigorous discussions and suggestions on how to check the reliability 
of the data; A. Greve, T.~P. Krichbaum, and J.~A. Zensus for technical
contributions to the global 86~GHz VLBP effort; and the CMVA group at
Haystack and V. Spinetti for their encouragement and inspiration during
the past two years.






\clearpage
\begin{deluxetable}{cccccccccc}
\tabletypesize{\scriptsize}
\tablecaption{Epoch $2000.30$ Modelfit Data for 3C\,273 and 3C\,279, 86~GHz\label{tbl-1}}
\tablewidth{0pt}
\tablehead{
\colhead{} & \colhead{} & \colhead{$r$} & \colhead{$\theta$} &
\colhead{$I$} & \colhead{$m$} & \colhead{$\chi$} &
\colhead{Major~Axis} & \colhead{Minor~Axis} & \colhead{$\phi$} \\
\colhead{Source} & \colhead{Component} & \colhead{(mas)} & \colhead{(deg)} &
\colhead{(Jy)} & \colhead{(\%)} & \colhead{(deg)} &
\colhead{(mas)} & \colhead{(mas)} & \colhead{(deg)} \\
\colhead{(1)} & \colhead{(2)} & \colhead{(3)} & \colhead{(4)} &
\colhead{(5)} & \colhead{(6)} & \colhead{(7)} &
\colhead{(8)} & \colhead{(9)} & \colhead{(10)}
}
\startdata
3C\,273 & D & $\ldots$ & $\ldots$ & $2.40$ & $< 1$ & $\ldots$ & $0.19$ & $< 0.05$ & $-4.3$ \\
& C2 & $1.06$ & $-123.4$ & $1.51$ & $11\pm3$ & $37\pm8$ & $0.37$ & $0.14$ & $4.5$ \\
& C1 & $2.09$ & $-130.8$ & $0.95$ & $< 1$ & $\ldots$ & $0.87$ & $0.39$ & $-17.5$ \\
& & & & & & & & & \\
3C\,279 & D & $\ldots$ & $\ldots$ & $10.23$ & $5\pm1$ & $-27\pm5$ & $0.15$ & $0.09$ & $21.2$ \\
& C1 & $0.94$ & $-122.6$ & $0.79$ & $20\pm7$ & $-26\pm10$ & $0.51$ & $< 0.05$ & $-26.5$ \\
\enddata
\tablecomments{Columns are as follows: (1) source name; (2) component name; (3) distance from
easternmost feature D; (4) position angle of separation from D; (5) total
intensity; (6) fractional linear polarization;
(7) uncalibrated orientation of the
linear polarization position angle; (8) major axis of the model component;
(9) minor axis of the model component; (10) orientation of the major axis.}
\end{deluxetable}


\clearpage


\begin{figure}
\figurenum{1}
\epsscale{0.49}
\plotone{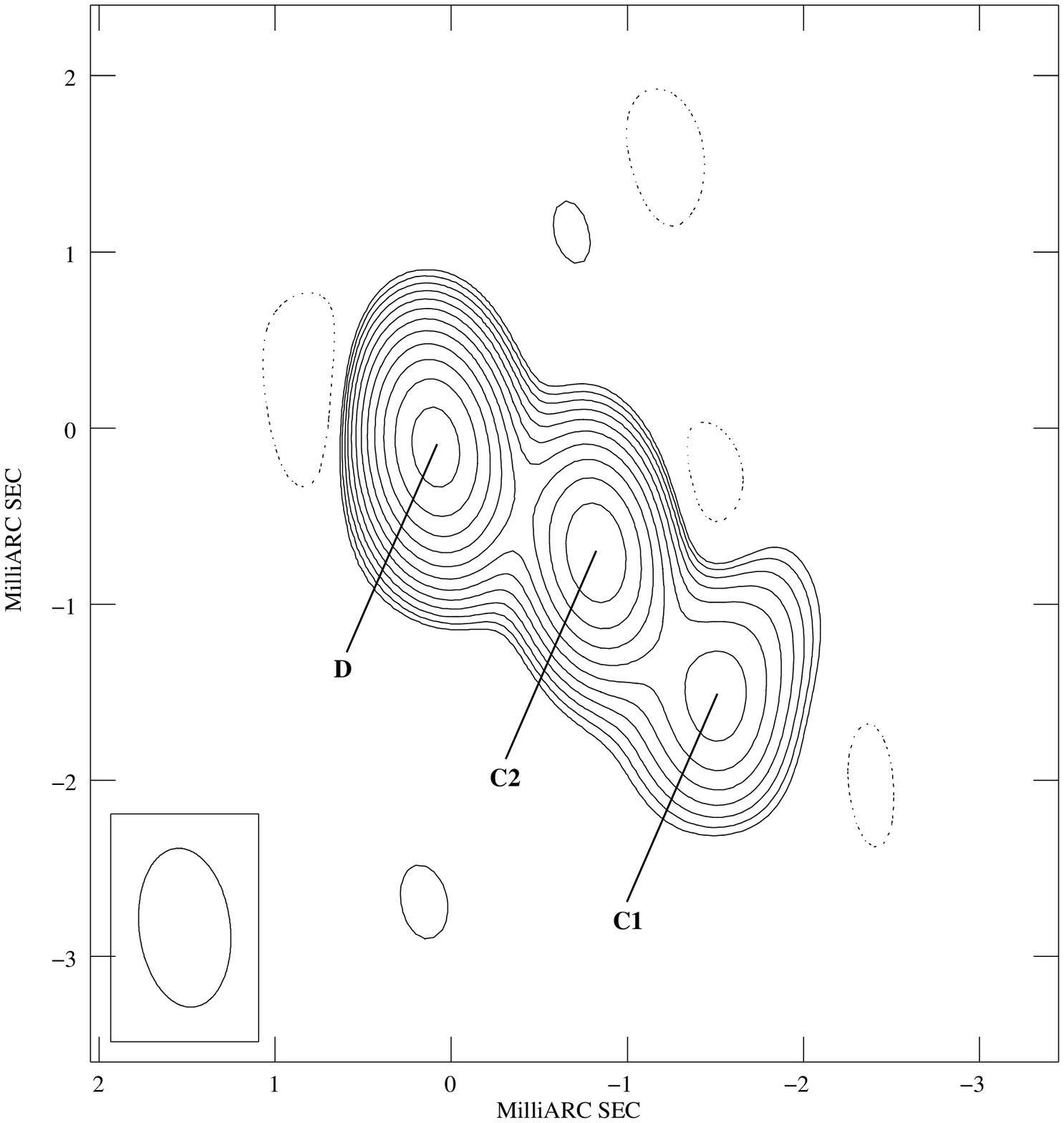}
\plotone{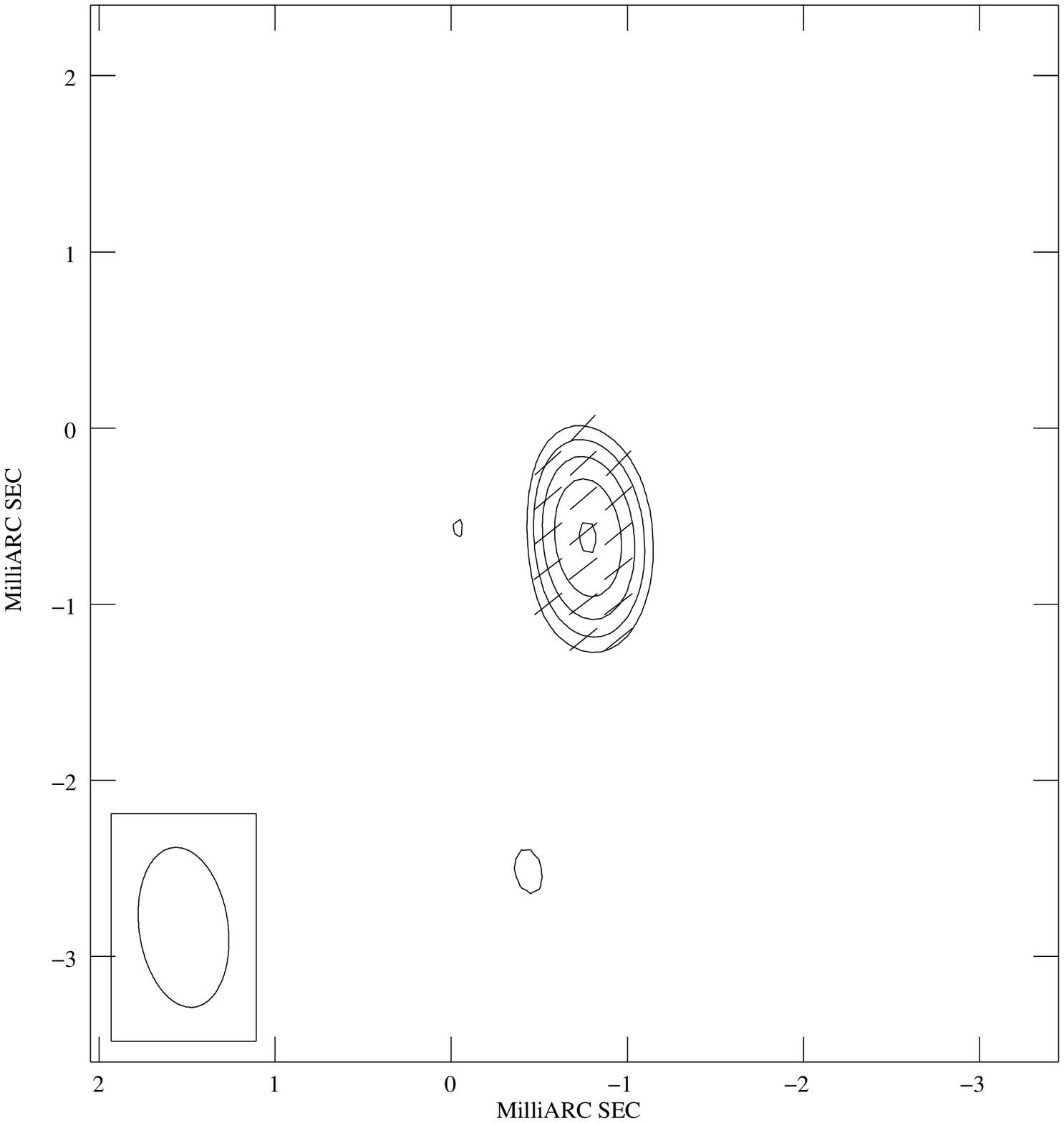}
\caption{Naturally weighted images of 3C\,273, 
epoch 2000.30, made with the CMVA at 86~GHz. ($a$) Total intensity distribution,
with contours of $I$ at $-40.00$, 40.00, 56.57, \ldots\ [factors of $\sqrt{2}$]
\ldots, 1280 and 1810 mJy~beam$^{-1}$; the peak is 2165~mJy~beam$^{-1}$ and
the restoring beam shown in the lower left is 0.91~mas$\times$~0.52~mas
at $\phi = 6.55^{\circ}$. ($b$) Linear polarization distribution, with 
contours of $p$ at 30.00, 42.43, \ldots\ [factors of $\sqrt{2}$] \ldots, 
84.85, and 120.0 mJy~beam$^{-1}$; the peak is 122.7~mJy~beam$^{-1}$. The sticks 
show the orientation $\chi$ of the electric field in the source rotated 
90$^{\circ}$ (see \S 4.2). The 
restoring beam is the same as for $I$. 
\label{fig1}}
\end{figure}

\begin{figure}
\figurenum{2}
\epsscale{0.49}
\plotone{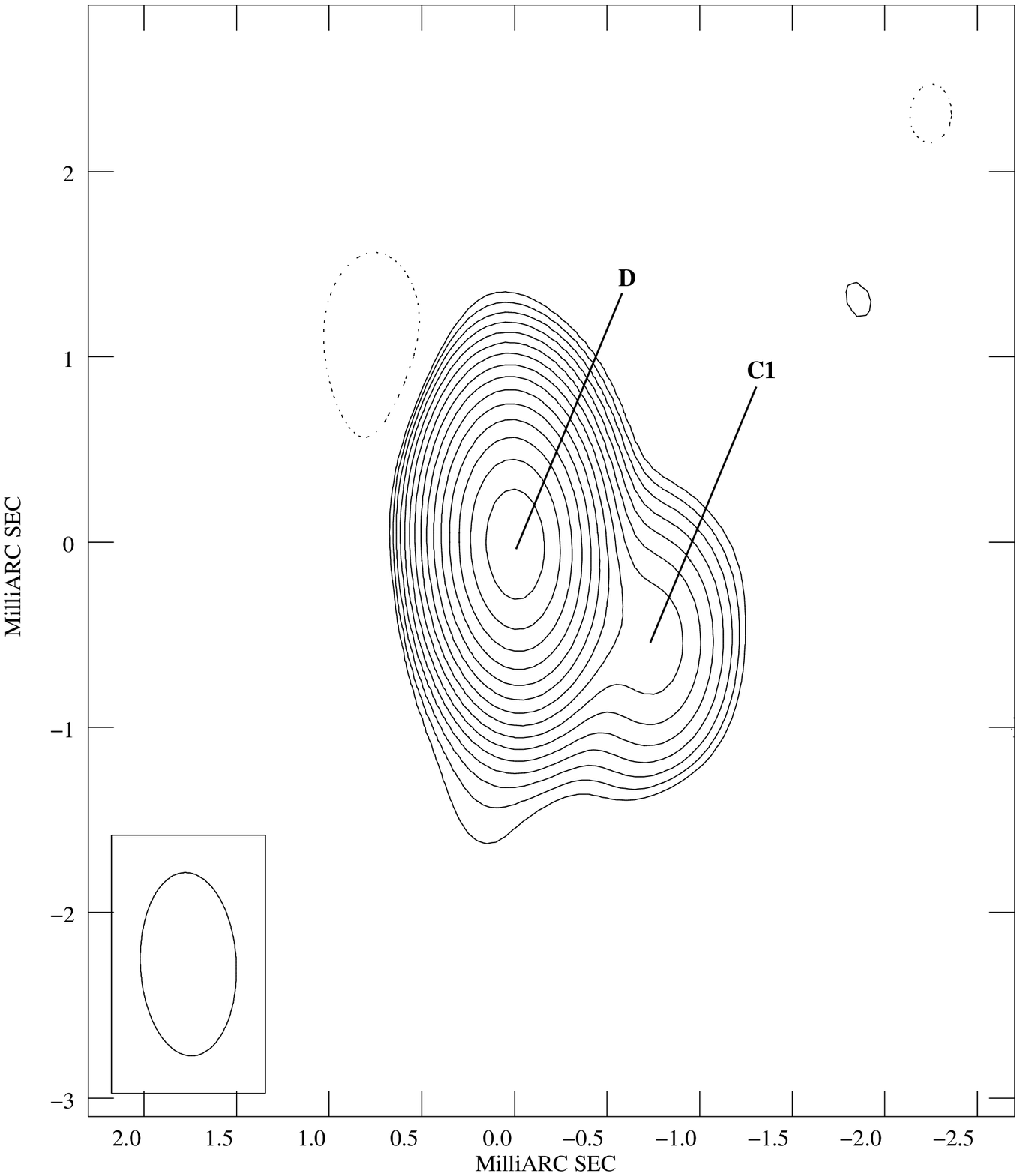}
\plotone{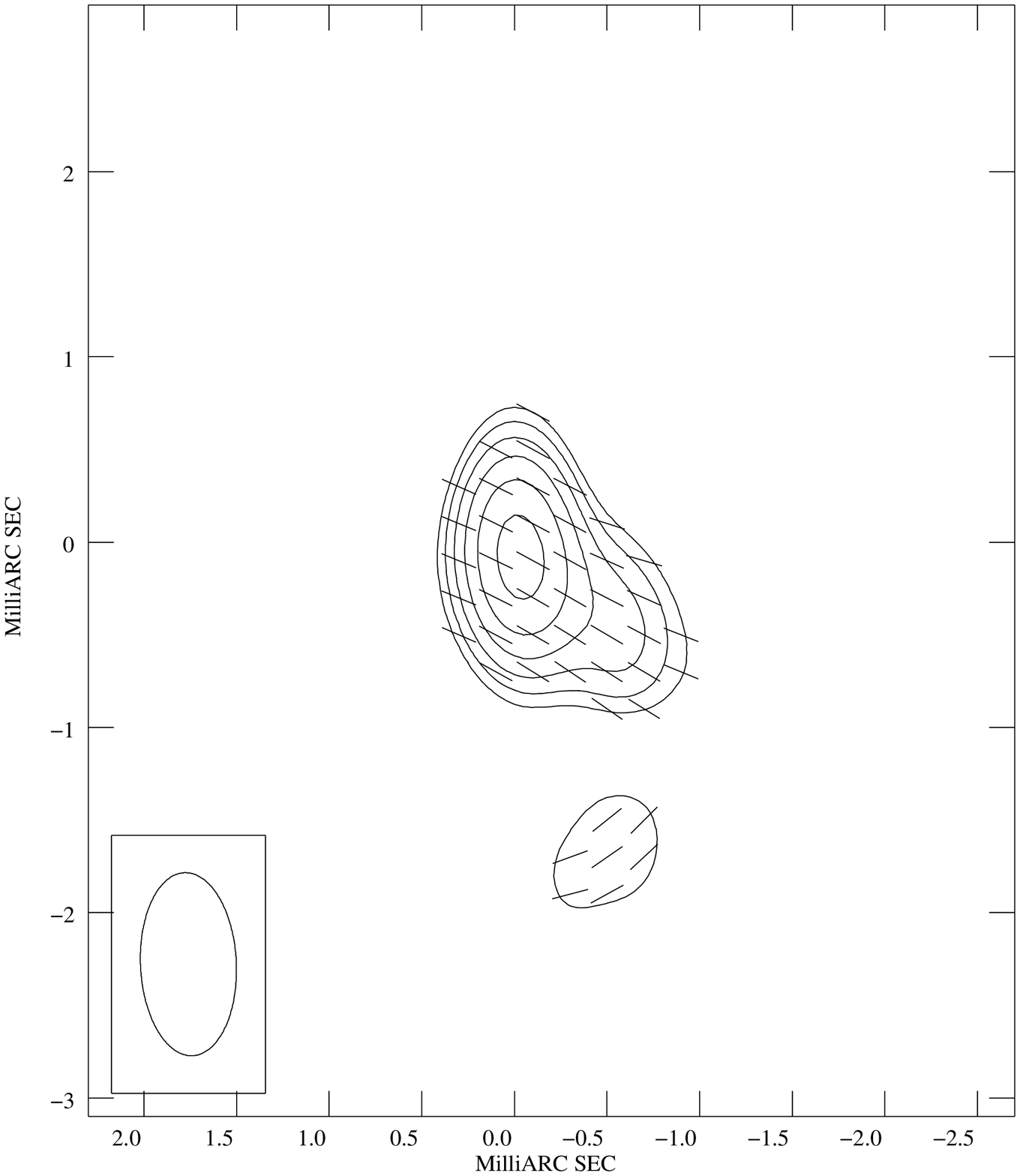}
\caption{Naturally weighted images of 3C\,279,                
epoch 2000.30, made with the CMVA at 86~GHz. ($a$) Total intensity distribution,
with contours of $I$ at $-60.00$, 60.00, 84.85, \ldots\ [factors of $\sqrt{2}$]
\ldots, 5431 and 7680 mJy~beam$^{-1}$; the peak is 9885~mJy~beam$^{-1}$ and
the restoring beam shown in the lower left is 0.99~mas$\times$~0.52~mas
at $\phi = 2.55^{\circ}$. ($b$) Linear polarization distribution, with 
contours of $p$ at 70.00, 98.99, \ldots\ [factors of $\sqrt{2}$] \ldots, 
280.0, and 396.0 mJy~beam$^{-1}$; the peak is 456.2~mJy~beam$^{-1}$. The sticks 
show the orientation $\chi$ of the electric field in the source rotated 
90$^{\circ}$ (see \S 4.2). The
restoring beam is the same as for $I$. 
\label{fig2}}
\end{figure}

\end{document}